\documentclass [amssymb, twocolumn, amsmath, showpacs]{revtex4-1}
\usepackage{graphicx,epsfig,amsfonts,amssymb}
\usepackage{bm}
\usepackage{times}
\begin{document}

\title{Landau-Zener Transitions in  Chains}

\author {N.~A. {Sinitsyn}$^{a}$ }


\address{$^a$ Theoretical Division, Los Alamos National Laboratory, Los Alamos, NM 87545,  USA}

\begin{abstract}
We determine transition probabilities in two exactly solvable
multistate Landau-Zener (LZ) models and discuss applications of our
results to the theory of dynamic passage through a  phase transition in the dissipationless quantum mechanical regime. In particular, we show that statistics
of particles in a new phase demonstrate  scaling behavior.  
Our results also reveal a symmetry that we claim is a property of a large class of multistate LZ models, whose explicit solutions are not presently known. We support our arguments by direct numerical simulations. 
\end{abstract}

\date{\today}

\maketitle

\section{Introduction}

The multistate Landau-Zener (LZ) problem is to determine transition amplitudes among $N$ discrete states after time evolution from $-\infty$ to $+\infty$ in systems described by the Sch\"odinger's equation with linearly time-dependent coefficients \cite{be}:

\begin{equation}
i\frac{d\psi}{dt} = (\hat{A} +\hat{B}t)\psi,
\label{mlz}
\end{equation} 
where $\hat{A}$ and $\hat{B}$ are constant $N\times N$ matrices. Any such a system can be transformed by a time-independent change of the basis  into its canonical form, in which the matrix $\hat{B}$ is diagonal and the off-diagonal matrix elements of  $\hat{A}$ are nonzero  only between eigenstates of $\hat{B}$ with different eigenvalues.  
Eigenstates of the matrix $\hat{B}$ are called the {\it diabatic} states. They transfer into eigenstates of the Hamiltonian when time is approaching $\pm \infty$. Therefore, for a finite number of coupled nondegenerate diabatic states, the scattering matrix can be defined. Arbitrary element $S_{nn'}$ of such an $N\times N$ matrix $\hat{S}$ is the amplitude of the diabatic state $n'$ at $t\rightarrow +\infty$, given that at $t\rightarrow -\infty$ the system was at the state $n$.  The related matrix $\hat{P}$, $P_{n\rightarrow n'}=|S_{nn'}|^2$, is called the matrix of transition probabilities.

Studies of multistate Landau-Zener models (MLZMs) were pioneered by
the article of Majorana \cite{maj} in which, in addition to an
independent discovery of the Landau-Zener formula \cite{LZ,maj}, he
showed that any solution of a spin-1/2  problem in time-dependent
magnetic fields, including the LZ-model, can be generalized to the
solution for dynamics of an arbitrary spin in the same magnetic
field. Hence, the Majorana's article \cite{maj} provided the
transition amplitudes in the MLZM with the Hamiltonian $\hat{H}=\beta
t \hat{S}_z+g \hat{S}_x$, where $\hat{S}$ is the arbitrary size spin operator.  Perhaps, the most unusual findings about the system (\ref{mlz}) are the so called Brundobler-Elser formula \cite{be} and the no-go theorem \cite{no-go} that provide expressions for some elements of the transition probability matrix in {\it any} model of the type (\ref{mlz}), suggesting that a strong progress in understanding complex MLZMs can be achieved.

Today, there is no general recipe to determine all transition amplitudes in  a complex MLZM analytically. In order to achieve this, one would have to consider  higher than 2nd order systems of differential equations with time-dependent coefficients. Nevertheless, 
a number of exact results \cite{be,vo1,vo2,no-go,be-comment,mlz7}, fully solvable nontrivial systems \cite{mlz0,mlz1,mlz2,mlz3,mlz4,mlz5,mlz6,mlz8,sinitsyn-02prb, sinitsyn-02prb2,mlz9}, and methods to study MLZMs \cite{sinitsyn-02prb2,str} of the type (\ref{mlz}) have been discovered. Some of the solved models were used in applications to condensed matter systems \cite{app}.

Here we add  two models to the class of the worked-out solvable MLZMs. 
Our models correspond to transitions on a semi-infinite chain of sites with a possibility of pair-wise jumps between neighboring sites. 
Specifically, we will provide transition probabilities for two models of type (\ref{mlz}), whose Schr\"odinger's equations for amplitudes read:
\vskip .3truecm
\underline {Model-1: Square-Root Growing Coupling}
\begin{equation}
i\dot{a}_n=-\beta n t a_n +\gamma \sqrt{n+1}a_{n+1} +\gamma \sqrt{n}a_{n-1}, \quad n \in N.
\label{h1}
\end{equation}

\underline {Model-2: Linearly Growing Coupling}
\begin{equation}
i\dot{a}_n=\beta n t a_n +\gamma(n+1)a_{n+1} +\gamma na_{n-1},  \quad n \in N,
\label{h2}
\end{equation}
with constant parameters $\beta$ and $\gamma$, and where $N$ is the set of nonnegative integer numbers.

The reason why both models are solvable is because their Hamiltonians are quadratic when they are written in terms of bosonic creation/annihilation operators, $\hat{a}$ and $\hat{a}^+$. Thus, if we identify states $|n\rangle$
with eigenstates of the boson number operator $\hat{a}^+\hat{a}$, then the Model-1 is  the Schr\"odinger's equation for state amplitudes with the Hamiltonian 
\begin{equation}
\hat{H}_1=-\beta t \hat{a}^+\hat{a} +\gamma(\hat{a}^+ + \hat{a}),
\label{hm1}
\end{equation}
and Model-2 is reproduced from the evolution of two coupled oscillators:
\begin{equation}
\hat{H}_2=\beta t \hat{a}^+\hat{a} +\gamma(\hat{a}^+\hat{b}^+ + \hat{a}\hat{b}),
\label{osc}
\end{equation}
under the condition that the initial population of the mode $\hat{a}$ is equal to the initial population of the mode $\hat{b}$.

Both models, (\ref{hm1}) and (\ref{osc}), have an important practical realization. They describe  a dynamic passage through a phase transition in a system of a molecular Bose condensate  interacting with cold atoms near the  Feshbach resonance. 
The Hamiltonian of this system is usually written as \cite{timm,yur,gur2}:

\begin{equation}
\hat{H}=\sum_{{\bm p},\sigma} \epsilon \hat{a}^+_{{\bm p},\sigma} \hat{a}_{{\bm p},\sigma} -\beta t \hat{b}^+\hat{b} + \frac{\gamma}{\sqrt{N}} \left[ 
\hat{b} \sum_{\bm p}  \hat{a}^+_{{\bm p},\uparrow}  \hat{a}^+_{{\bm p},\downarrow} +{\rm h.c.}
\right],
\label{gh}
\end{equation}
where $\hat{b}$ is the diatomic molecule annihilation operator and $ \hat{a}_{{\bm p},\sigma}$ is the annihilation operator of a single atom with momentum 
${\bm p}$ and spin $\sigma$; $N$ is the number of atoms and $\gamma$ describes the strength of the conversion of atoms into molecules near the resonance.
Model-1 appears in the limit when atoms are in a macroscopic condensate state  
 so that their operators can be treated as constant c-numbers. It describes creation of a molecular condensate beyond the mean-field assumption for the molecular field $\hat{b}$. In particular, it can be applied to the experimentally important case with zero initial number of molecules. 
 Model-2 corresponds to the opposite process of a decay of a molecular condensate into atomic modes. In this case, one assumes that the system  has initially macroscopic number of Bose condensed molecules that pass through the Feshbach
 resonance and split into pairs of atoms with initially close to zero populations. 
  
 Both models (\ref{hm1}) and (\ref{osc}) have been studied previously for application to the transition through the Feshbach resonance \cite{abanov,gur2,sinitsyn-pra}, however, the focus was either on the average number of molecules/atoms converted during the process or on the evolution from the ground state. 
 In this work we will explore the exact statistics of the number of defects (i.e particles in a new phase) created during the evolution starting from an {\it arbitrary} initial state, and stressing universality and scaling in the full probability distribution of the possible outcomes of the sweeping through a phase transition process.

The structure of our article is as follows. In Sections II and III, we derive the state-to-state transition probabilities in, respectively, Model-1  and Model-2.  In Section IV, we  discuss possible applications of our results in the theory of dynamic quantum phase transitions.
Appendix A  is devoted to the numerical study of the symmetry of the transition probability matrix
that we initially observed in  solutions of our models and then claimed that it is actually the property of all MLZMs in linear chains. We summarize our results in Conclusion.

\section{Transition probabilities in Model-1}  

The Hamiltonian of the Model-1 can be easily transformed into the Hamiltonian of a quantum harmonic oscillator with a time-dependent force. Such systems were very well studied previously for various applications. For example, one can engineer the Hamitonian of such a system in an array of optical couplers and, literarily, observe the evolution of transition amplitudes with ``time" \cite{couplers}.
   
To simplify our notation, first, we reduce the number of parameters by  rescaling: $t \rightarrow t/\sqrt{\beta}$, $g = \gamma/\sqrt{\beta}$. 
Next, we remove strongly oscillating phase factors at $t\rightarrow \pm \infty$ by changing the basis
\begin{equation}
a_n(t) \rightarrow a_n(t) e^{i nt^2/2}, 
\label{gauge}
\end{equation}
which  does not change the transition probabilities. After these transformations, the Hamiltonian (\ref{hm1}) is
\begin{equation}
\hat{H}_1=ge^{-it^2/2} \hat{a}^+ + ge^{it^2/2} \hat{a}.
\label{hm12}
\end{equation}
We will search for the solution of the corresponding  Schr\"odinger's equation in the form of a coherent state ansatz:
\begin{equation}
\Psi(t) = e^{-\frac{|\alpha_{-\infty}|^2}{2}} e^{\phi(t) + \alpha(t) \hat{a}^+ } |0\rangle. 
\label{coh1}
\end{equation} 
Substituting this vector in equation $i\dot{\Psi} = \hat{H}_1 \Psi$ and collecting separately c-terms near $\Psi(t)$ and $\hat{a}^+ \Psi(t)$ we find a pair of equations:

\begin{equation}
i\dot{\phi}=ge^{it^2/2} \alpha, \quad i\dot{\alpha}=ge^{-it^2/2}.
\label{coh2}
\end{equation}
With initial conditions $ \alpha( -\infty )= \alpha_{-\infty}$ and $\phi( -\infty) = 0$, they have solutions
\begin{eqnarray}
\alpha(t) &=& \alpha_{-\infty} -ig\int_{-\infty}^t dt_1e^{-it_1^2/2} , \\
\\
\phi(t) &=& -ig\alpha_{-\infty} \int_{-\infty}^t dt_1e^{it_1^2/2} - \\
&-&g^2 \int_{-\infty}^t dt_1 \int_{-\infty}^{t_1} dt_2 e^{it_1^2/2 -it_2^2/2}.
\label{coh3}
\end{eqnarray}

For the time-evolution from  minus to plus infinity, this gives us $\alpha(+\infty) = \alpha_{-\infty} - g\sqrt{2\pi}e^{i\pi/4}$ and $\phi(+\infty) = - g^2\pi+g\alpha_{-\infty}\sqrt{2\pi}e^{-i\pi/4}$. Hence, if the initial state is
a coherent state, $\Psi_{t\rightarrow - \infty}= | \alpha_{-\infty} \rangle$, the state after the transition becomes

\begin{eqnarray}
\Psi_{+ \infty}= e^{-\frac{|\alpha_{-\infty}|^2}{2}- g^2\pi+g\alpha_{-\infty}\sqrt{2\pi}e^{-i\pi/4}} \times \\
\times e^{  (\alpha_{-\infty} - g\sqrt{2\pi}e^{i\pi/4}) \hat{a}^+ } |0\rangle.
\label{coh44} 
\end{eqnarray}

A specific case of an initial state without defects: $\Psi_{- \infty}=|0 \rangle$ corresponds to $\alpha_{-\infty}=0$, and Eq. (\ref{coh44}) gives the wave function in the form
\begin{equation}
\Psi_{+ \infty}^0= e^{- g^2\pi} e^{- g\sqrt{2\pi}e^{i\pi/4} \hat{a}^+ } |0\rangle,
\label{coh4}
\end{equation}
which is a coherent state with $\alpha=-g\sqrt{2\pi}e^{i\pi/4}$. From $e^{\alpha \hat{a}^+ } |0\rangle = \sum_{n=0}^{\infty} (\alpha^n)/\sqrt{n!} |n\rangle$, we can explicitly read the transition amplitudes between states $0$ and $n$ as
$S_{0n}= e^{- g^2\pi} (-g\sqrt{2\pi})^n e^{in\pi/4}/\sqrt{n!}$, and transition probabilities are explicitly given by
\begin{equation}
P_{0 \rightarrow n}\equiv |S_{0n}|^2 =  \frac{e^{- 2\pi g^2} (2\pi g^2)^n}{n!}.
\label{p00n}
\end{equation}
The number of defects (i.e. molecules in the created molecular condensate) has the Poisson distribution (\ref{p00n}) with the average number  $\langle n \rangle = 2\pi g^2$. Returning to the parametrization (\ref{hm1}), we find $\langle n \rangle =2\pi \gamma^2/\beta$, i.e. the average number of defects scales as $\beta^{-1}$ with the sweeping rate. Two additional consequences follow from the knowledge of the full distribution (\ref{p00n}). First, the Poisson distribution has all equal cumulants, so that this scaling is valid not only for the average number of defects but also for their variance, skewness and so on. The second observation is that the probability of not creating any defect  depends exponentially on the inverse sweeping rate $\beta$: 
\begin{equation}
P_{0 \rightarrow 0}=e^{- 2\pi \gamma^2/\beta} .
\label{p00}
\end{equation}
Later, we will return to this observation and claim that this behavior is, in fact, more general than the scaling of the number of defects with the sweeping rate through a phase transition.

In order to determine transition probabilities between any pair of levels $n$ and $n'$, we use the fact that any state $|n\rangle$ can be written as a superposition of coherent states $|e^{i\theta} \rangle$ as

\begin{equation}
|n\rangle = \frac{e^{1/2}}{2\pi} \sqrt{n!}\int_{0}^{2\pi}d\theta e^{-in\theta} \vert e^{i\theta} \rangle.
\label{nn}
\end{equation}
Combining this with (\ref{coh4}) we find that the state vector that starts as $|n\rangle$ transforms into

\begin{eqnarray}
\Psi_{+ \infty}&=& \frac{\sqrt{n!} e^{- g^2\pi}}{2\pi} \int_{0}^{2\pi}d\theta e^{-in\theta} \times \\
\nonumber &\times& {\rm exp} \left( g \sqrt{2\pi}e^{-i\pi/4}e^{i\theta}+ 
(e^{i\theta} - g\sqrt{2\pi}e^{i\pi/4}) \hat{a}^+ \right) \vert 0\rangle.
\label{coh6}
\end{eqnarray}
For the transition amplitude to the state $|n'\rangle$ with $n'\geq n$ we find
\begin{eqnarray}
n'\geq n&:& \,\, S_{nn'}=e^{-\pi g^2} (-\sqrt{2\pi}ge^{i\pi/4})^{n'-n}\sqrt{\frac{n!}{n'!}}  \\
\nonumber &\times& \sum_{m=0}^n \frac{n'! }{m!(n-m)!(n'-n+m)!} (-2\pi g^2)^m.
\label{pnn1}
\end{eqnarray}
Respectively, for $n'<n$ we find 
\begin{eqnarray}
\nonumber &n&'<n: \quad S_{nn'}=e^{-\pi g^2} (-\sqrt{2\pi}g)^{n'+n}e^{i\pi(n'-n)/4}  \times \\
&\times& \sqrt{\frac{n!}{n'!}}\sum_{m=0}^{n'} \frac{n'! (-2\pi g^2)^{-m}}{m!(n'-m)!(n-m)!} .
\label{pnn22}
\end{eqnarray}

\begin{figure}[t]
\includegraphics[width=3.4in]{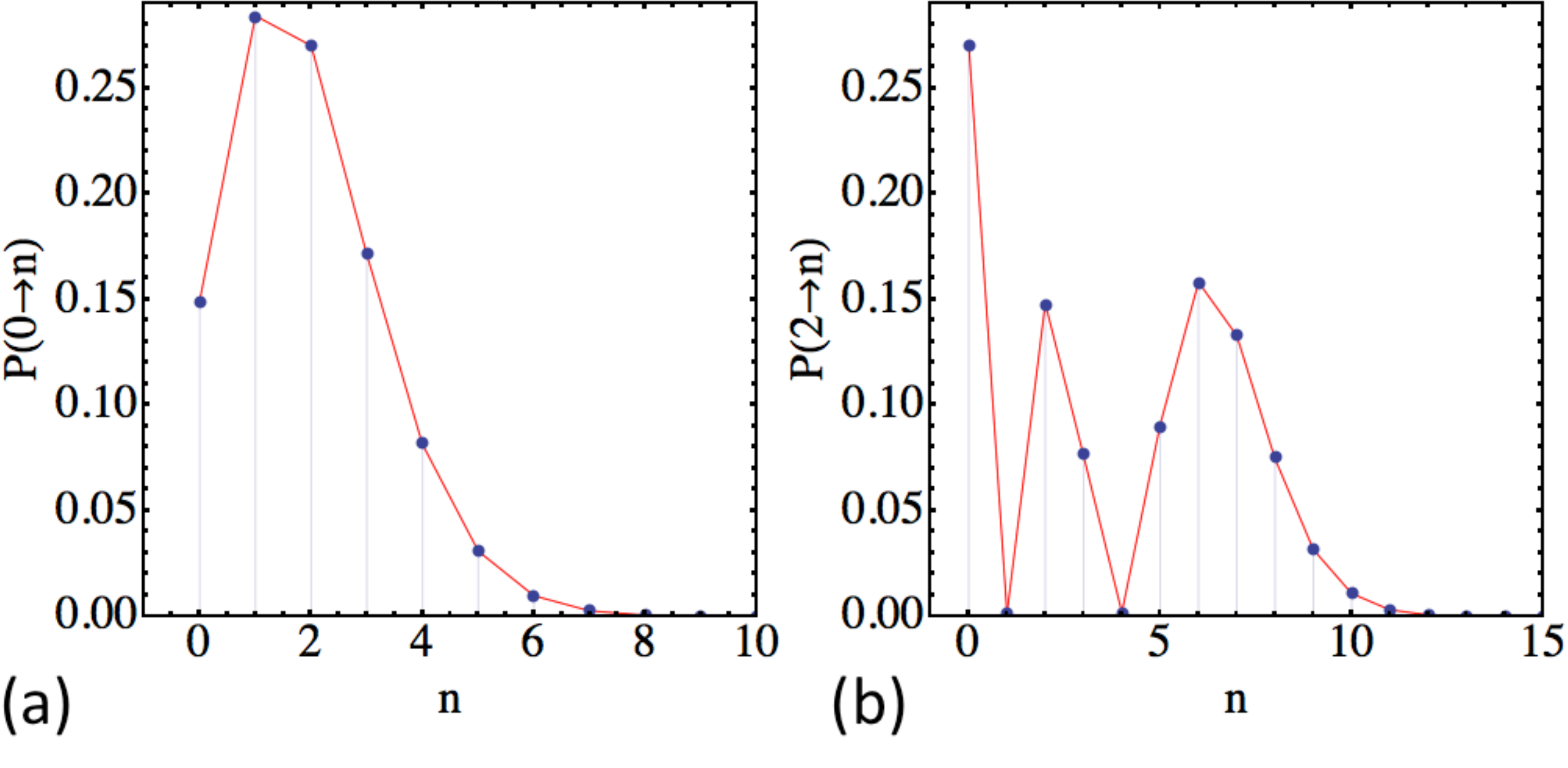}
 \caption{(Color online) The numerical check (blue dots) of the theoretical prediction (the red curve connecting discrete points) based on Eq. (\ref{probs}) for (a) transition probabilities from the state $0$ to the state $n$ at $g=0.55$; 
 (b) transition probabilities from the state $2$ to the state $n$ at $g=0.55$. Time evolution is from $t=-600$ to $t=600$.}
 \label{m1-check}
 \end{figure}

Transition probabilities $P_{n\rightarrow n'}=|S_{nn'}|^2$ can be compactly written in terms of the known special functions:
\begin{eqnarray}
\label{probs}
\nonumber n'>n: \quad P_{n \rightarrow n'}&=&\frac{e^{-2\pi g^2}n!(2\pi g^2)^{n'-n}}{n'!} \left( L_n^{n'-n} (2\pi g^2) \right)^2,\\
n'=n: \quad P_{n \rightarrow n}&=&e^{-2\pi g^2} \left( L_n (2\pi g^2) \right)^2,\\
\nonumber n'<n: \quad P_{n \rightarrow n'}&=&\frac{e^{-2\pi g^2}n'!(2\pi g^2)^{n-n'}}{n!} \left( L_{n'}^{n-n'} (2\pi g^2) \right)^2,
\end{eqnarray}
where $L_n$ is the n-th Laguerre polynomial and $L_n^k$ is the associated Laguerre polynomial. 
In Fig.~\ref{m1-check} we compare predictions (\ref{probs})  with our direct numerical simulations of (\ref{h1}). Results appear to be in a perfect agreement with each other.

Interestingly, amplitudes of transitions $S_{nn'}$ and $S_{n'n}$ are different only by a phase factor, which corresponds to the symmetry of the transition probability matrix: 
\begin{equation}
P_{n\rightarrow n'} =P_{n' \rightarrow n}. 
\label{psn}
\end{equation}
This fact is quite surprising considering that there seems to be no obvious symmetry in the evolution equation of the model that would lead to (\ref{psn}). In fact, our numerical simulations show that this property is {\it generic} i.e. it is found in all MLZMs in finite and infinite linear chains. We will discuss this property in more detail in Appendix. 

Some transition probabilities with lowest $n$ and $n'$  are explicitly given by 
\begin{eqnarray}
\nonumber &P&_{1 \rightarrow 1}= e^{- 2\pi g^2} (1-2\pi g^2)^2, \\
\nonumber &P&_{1 \rightarrow 2}=P_{2 \rightarrow 1}=e^{- 2\pi g^2} 4\pi g^2 (g^2\pi-1)^2,\\
&P&_{2\rightarrow 2}=e^{- 2\pi g^2} (1-4\pi g^2+2\pi^2g^4)^2,\\
\nonumber &P&_{1\rightarrow 3}=P_{3\rightarrow1}=e^{- 2\pi g^2} \frac{2}{3} \pi^2 g^4(3-2\pi g^2)^2, \\
\nonumber &P&_{2\rightarrow 3}=P_{3\rightarrow 2}=e^{- 2\pi g^2}\frac{2}{3}\pi g^2(3-6\pi g^2+2\pi^2g^4)^2,\\
\nonumber &P&_{3\rightarrow 3}=e^{- 2\pi g^2}\frac{1}{9}(3- 18\pi g^2   +18\pi^2g^4-4\pi^3 g^6)^2.
\label{pother}
\end{eqnarray}
It is useful  to compare this list of transition probabilities with the ones for other known MLZMs. Thus, in all known exact solutions with a finite number of states, e.g. \cite{maj, mlz1,mlz2}, transition probabilities can be expressed as finite polynomials  of exponents ${\rm exp}(-\pi g_{ij}^2/\beta_{ij})$, where $g_{ij}$ and $\beta_{ij}$ are, respectively, off-diagonal couplings and slope differences between  pairs of diabatic states $i$ and $j$. In another extreme, the exact solution of the MLZM in an infinite chain with constant couplings \cite{sinitsyn-02prb} returns transition probabilities in terms of the Bessel function: $P_{n \rightarrow n'}=J_{|n-n'|}^2(\sqrt{8\pi g^2})$. Our Model-1 shows features of both these classes. Transition probabilities contain an exponent $e^{- 2\pi g^2}$ but, unlike the known solved finite size MLZMs, this exponent is multiplied by polynomials  of $\pi g^2$ rather than exponents of this combination.

\section{Transition Probabilities in Model-2}  

In order to solve  Model-2, we adopt a different strategy \cite{sinitsyn-02prb}. 
Let us introduce the amplitude generating  function $u=u(z,t) = \sum_n a_n z^n$.
 After rescaling, $t \rightarrow t/\sqrt{\beta}$ and $g=\gamma/\sqrt{\beta}$, Eq. (\ref{h2}) becomes  a linear partial differential equation in terms of $u(z,t)$:
\begin{equation}
\partial_t u+i[ z t +g(z^2+1)] \partial_{z} u = -igzu,
\label{u1}
\end{equation}
which can be solved by the method of characteristics, leading to equations:
\begin{equation}
\frac{d u}{dt} = -igzu,
\label{c1}
\end{equation}
\begin{equation}
\frac{d z}{dt} = i[zt+g(z^2+1)].
\label{c2}
\end{equation}

Nonlinear Eq. (\ref{c2}) can be transformed into the linear 2nd order differential equation by a change of variables
\begin{equation}
z(t)=\frac{i\partial_t a(t)}{ga(t)},
\label{aa}
\end{equation}
where
\begin{equation}
a''-ita'-g^2a=0.
\label{aa1}
\end{equation}
One of the initial conditions in (\ref{aa1}) can be chosen arbitrarily. We will assume that $|a(-\infty)|=1$. The second initial condition is given by 
\begin{equation}
z_{-\infty}=a'(-\infty)/a(-\infty),
\label{as2}
\end{equation}
where $z_{-\infty}$ is the value of $z(t)$ at $t \rightarrow - \infty$.
Substituting (\ref{aa}) into (\ref{c1}) we also find
\begin{equation}
u(t)=u_{-\infty} a(t).
\label{u2}
\end{equation}

Equation (\ref{aa1}) has two solutions with leading asymptotics at $t \rightarrow -\infty$:
\begin{equation}
a_1(-\infty)  \sim t^{ig^2}, \quad a_2(-\infty) \sim -i t^{-ig^2-1} e^{it^2/2}.
\label{as}
\end{equation}
The Wronskian of such two solutions is equal to unity: $W=a_1(t)a_2'(t)-a_2(t)a_1'(t)=1$. 
Substituting (\ref{as}) into (\ref{as2}) we find that 

\begin{equation}
a(t)=a_1(t)+z_{-\infty} (g/i)a_2(t).
\label{aa3}
\end{equation}
Hence,
\begin{equation}
z(t)=\frac{a_1'(t)+z_{-\infty}(g/i)a_2'(t)}{a_1(t)+z_{-\infty}(g/i)a_2(t)}\frac{i}{g}.
\label{aa4}
\end{equation}
In particular, this allows us to express $z_{-\infty}$ in terms of $z_{+\infty}$ that is the value of $z(t)$ at $+\infty$:

\begin{equation}
z_{-\infty}=\frac{a_1(+\infty) z_{+\infty}(g/i)-a_1'(+\infty)}{a_2'(+\infty)-z_{+\infty}(g/i)a_2(+\infty)}\frac{i}{g}.
\label{zz}
\end{equation}

Assuming that the system starts on the level $|n\rangle$, we have $u_{-\infty} =z_{-\infty}^n$.
Substituting this into (\ref{u2}) we find 
\begin{equation}
u(t)=z_{-\infty}^n [a_1(t)+z_{-\infty}(g/i)a_2(t)].
\label{u3}
\end{equation}
From (\ref{u3}) and (\ref{zz}),  we obtain for $z\equiv z_{+\infty}$:

\begin{eqnarray}
 u(z)=\left(\frac{a_1(+\infty) z(g/i)-a_1'(+\infty)}{a_2'(+\infty)-z(g/i)a_2(+\infty)}\frac{i}{g} \right)^n \times \\
\nonumber \times \frac{1}{a_2'(+\infty)-z(g/i)a_2(+\infty)}.
\label{uqq5} 
\end{eqnarray}
\begin{figure}[t]
\includegraphics[width=3.4in]{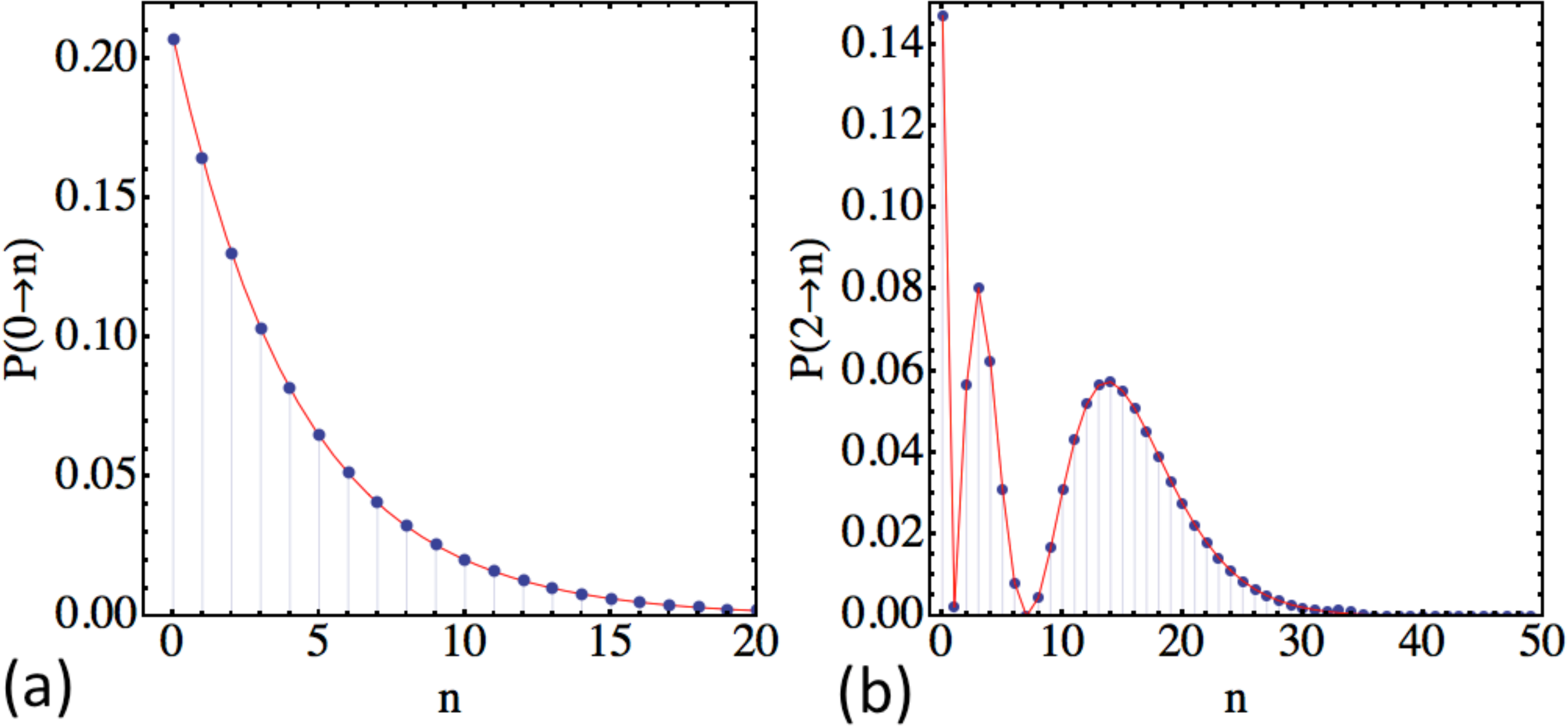}
 \caption{(Color online) The numerical check (blue dots) of the theoretical prediction (red curves) based on Eq. (\ref{u4}) for (a) transition probabilities from the state $0$ to the state $n$ at $g=0.5$; (b) transition probabilities from the state $2$ to the state $n$ at $g=0.4$. Time in numerical simulations changes from $t=-700$ to $t=700$.}
 \label{m2-check}
 \end{figure}
Asymptotics of (\ref{aa1}) at $t \rightarrow +\infty$ are known \cite{yur}. Their non-vanishing contributions read:
\begin{eqnarray}
\nonumber a_1(+\infty) &\sim& e^{\pi g^2}   t^{ig^2}, \\
\nonumber a_1'(+\infty) &\sim& g \sqrt{e^{2\pi g^2}-1} e^{-i\pi/4} e^{it^2/2} t^{-ig^2+i{\rm arg}\Gamma (ig^2)},\\
\nonumber a_2(+\infty) &\sim&  t^{ig^2}e^{-i{\rm arg}\Gamma (ig^2)}e^{-i\pi/4} \sqrt{e^{2\pi g^2}-1}/g,\\
 a_2'(+\infty) &\sim& -ie^{\pi g^2} e^{it^2/2} t^{-ig^2}. 
\label{as4}
\end{eqnarray}

 It is possible to simplify the generating function $u(z)$ by noticing that the multiplication of $z$ by any complex number with a unit absolute value is equivalent to a mere phase transformation of all amplitudes $a_n$, which does not influence the transition probabilities. Similarly, multiplication of $u$ by an arbitrary phase factor, only shifts phases of $a_n$ but not their absolute values. 
The time-dependent factors, $e^{it^2/2} t^{ig^2}$, can be moved in front of $u$ and the remaining time-dependent phase can be absorbed by $z$. Same can be applied to all factors $e^{\pm i{\rm arg}\Gamma (ig^2)}$, $i$, $1/i$ and 
$e^{i\pi/4}$ in  (\ref{as4}). The remaining trimmed generating function of transition amplitudes from the state $|n\rangle$ to any other state then explicitly reads:

\begin{equation}
u(z)= \left(\frac{ z-\sqrt{1-e^{-2\pi g^2}}}{1-z\sqrt{1-e^{-2\pi g^2}}} \right)^n  \frac{e^{-\pi g^2}}{1-z\sqrt{1-e^{-2\pi g^2}}}.
\label{u4}
\end{equation}

In particular, by setting $z=0$ in (\ref{u4}) and taking the square of the result we find the probability of a transition from the state $|n\rangle $ into the empty state $|0\rangle$:
 \begin{equation}
P_{n\rightarrow 0} = e^{-2\pi g^2} \left(1-e^{-2\pi g^2} \right)^n,
\label{pn0}
\end{equation}
which is the geometric distribution. From the previous studies of this model \cite{sinitsyn-pra} we know that each defect that was present at the initial state creates, on average, an exponentially large number of new defects, $\langle n (+\infty) \rangle \sim n(-\infty) e^{2\pi g^2}$ . Equation (\ref{pn0}) shows that despite this tendency the transition probability into the empty state decays relatively slowly with the initial number of defects.  Also, the result appears to be similar to the known solutions of MLZMs with a finite number of states in the sense that 
it depends only on the simple powers of $e^{-2\pi g^2}$.

According to (\ref{pn0}), the probability of creating no defects if the evolution starts at the ground state is given by a simple exponent:
\begin{equation}
P_{0\rightarrow 0} = e^{-2\pi \gamma^2/\beta},
\label{pp0}
\end{equation}
where we again recalled that $g=\gamma/ \sqrt{\beta}$ in order to highlight the exponential dependence of $P_{00}$ on the inverse sweeping rate $1/\beta$, which is usually a directly controlled parameter in experiments.

If the evolution starts with the ground state $|0\rangle$ then
\begin{equation}
u_0(z)=\frac{1}{e^{\pi g^2}-z\sqrt{e^{2\pi g^2}-1}}.
\label{uu4}
\end{equation}
Expanding (\ref{uu4})  as a Taylor series  we obtain the individual transition amplitudes. Taking their absolute square values we obtain transition probabilities from $0$  to all states in a closed form:
\begin{equation}
P_{0\rightarrow n} = e^{-2\pi g^2} \left(1-e^{-2\pi g^2} \right)^n.
\label{ppn}
\end{equation}
This particular distribution (\ref{ppn}), but not the full amplitude generating function (\ref{u4}), was previously found in \cite{gur2} by a different approach. Comparing (\ref{ppn}) and (\ref{pn0}) we again find the symmetry $P_{n\rightarrow 0}=P_{0\rightarrow n}$.

Figure~\ref{m2-check} shows a perfect agreement of the analytical prediction of (\ref{u4}) with our numerical simulations. Below, we also provide some explicit transition probabilities with lowest initial and final state indexes:  
\begin{eqnarray}
\nonumber &P&_{1\rightarrow 1}= e^{- 2\pi g^2} (1-2e^{- 2\pi g^2})^2, \\
\nonumber &P&_{1\rightarrow 2}=P_{2 \rightarrow 1}=e^{- 2\pi g^2} (1-e^{- 2\pi g^2})(1-3e^{- 2\pi g^2}))^2,\\
&P&_{2\rightarrow 2}=e^{- 2\pi g^2} (1-6e^{- 2\pi g^2}+6e^{- 4\pi g^2})^2,\\
\nonumber &P&_{1\rightarrow 3}=P_{3\rightarrow 1}=e^{- 2\pi g^2} (1-5e^{- 2\pi g^2}+4e^{- 4\pi g^2})^2, \\
\nonumber &P&_{2\rightarrow 3}=P_{3\rightarrow 2}=e^{- 2\pi g^2}(1-e^{- 2\pi g^2})(1+\\
\nonumber &+&2e^{- 2\pi g^2}(5e^{- 2\pi g^2}-4))^2,\\
\nonumber &P&_{3\rightarrow 3}=e^{- 2\pi g^2}(1-2e^{- 2\pi g^2})^2(1-10e^{- 2\pi g^2}(1-e^{- 2\pi g^2}))^2.
\label{pother}
\end{eqnarray}
Again, we always find the symmetry of the transition probability matrix: $P_{n\rightarrow n'}=P_{n' \rightarrow n}$.

\section{Application to dynamic phase transitions}

Modern studies of non-adiabatic effects during a passage through a quantum phase transition often concentrate on the almost adiabatic regime, in which a system almost reaches the new ground state after the
sweep of a control parameter through a phase transition \cite{zurek}. Nonadiabatic corrections in such a process 
are substantial only for small energy excitations defined by the sweeping rate $\beta$. For systems with a continuous spectrum they result in scaling of the number of excitations (defects) with $\beta$.
Such a nearly adiabatic regime is usually hard to achieve in  macroscopic and even mesoscopic quantum systems. There are many experimentally reported counterexamples to the naive application of the Landau-Zener formula to 
mesoscopic/macroscopic systems. Thus, an adiabatic sweep of a magnetic field
through the paramagnetic phase of an ensemble of initially polarized dipole-coupled nano-magnets usually leads to  demagnetization rather than the transition to the ground state with the opposite polarization \cite{nanomag}. 
Similarly, the conversion efficiency of an atomic condensate into molecules after the passage through the Feshbach resonance is often close to $1/2$ rather than $1$ at the lowest temperatures  in the adiabatic limit \cite{fesh,dobrescu}. 
Such deviations from the Landau-Zener theory have been explained, in particular, by the sensitivity of the adiabatic limit of the Landau-Zener formula to various decoherence effects \cite{LZ-noise}, which are generally 
present in mesoscopic and macroscopic systems, and by the breaking of the adiabatic approximation by many-body interactions \cite{gur1,gur2}. 
 
The models that we solved in previous sections correspond to a different regime of a relatively fast sweep throughout a new phase. In this regime, the initial phase is  mostly preserved during the process, however, the number of defects (molecules in Model-1 and atomic pairs in Model-2) can be  large so that collective effects in their dynamics play the role. 
Such a regime is much less vulnerable to decoherence \cite{garanin} due to much smaller time allowed for decoherence effects to accumulate, as well as due to specific robustness of the Landau-Zener processes to noise effects in this limit \cite{spin-bath}.
 
We found that the probability distributions of the number of defects
in this regime can be distinctly nonclassical. Consider, for example, a classical stochastic model of Markovian independent particle creation with a constant rate. The number of particles  in such a process has the Poisson distribution with the mean $\langle n \rangle$ and an exponentially small probability of creating zero number of particles $P_{0\rightarrow 0} \sim e^{-\langle n \rangle}$.
In contrast, for Model-2, it has been shown in \cite{sinitsyn-pra}
that if the passage through the phase transition starts with no
defects, then $\langle n \rangle \sim e^{2\pi g^2}$, while we showed
here that $P_{0\rightarrow 0} = e^{-2\pi g^2}$, i.e. $P_{0\rightarrow 0} \sim \langle n
\rangle^{-1}$, which is dramatically different from the stochastic
particle creation case. For example, if on average we observe 1000
defects, then it would be sufficient to repeat the experiment $\sim
1000$ times to observe an event without a defect creation, while it
would take $\sim e^{1000}$ trials to observe such an event if
particles were generated by a classical stochastic process without
memory.  Our exact result shows that this property of the enhanced
transition probability to the empty state persists even when the
initial state is already populated with a number of defects.  

In studies of the quantum mechanical regime of a  fast passage through a phase transition, the probability to create no defects should play a special role also for the reason of its universal scaling with the inverse sweeping rate $1/\beta$.  
Eqs. (\ref{p00}) and (\ref{pp0}) show that the probability $P_{0\rightarrow 0}$ is the same in two models with very different behavior of other transition probabilities. This fact is not a coincidence but rather a direct consequence of the Brundobler-Elser formula \cite{be},
which was initially found in numerical simulations \cite{be} and which by now is a mathematically rigorously  proved result in the multi-state LZ theory \cite{vo1,be-comment}. According to this formula, if the system starts its evolution from the ground state and evolves according to Eq. (\ref{mlz}) in time from $-\infty$ to $+\infty$, then the probability to remain in the same diabatic state is given by
\begin{equation}
P_{0\rightarrow 0}={\rm exp} \left( -2\pi \sum_{i=1}^N \frac{|\gamma_{0i}|^2}{\beta_0-\beta_i }\right),
\label{be}
\end{equation} 
where $\gamma_{0i}$ is the coupling between the state $|0\rangle$ and the excitation state $|i \rangle$, and where $\beta_0$ and $\beta_i$ are the slopes of diabatic levels of these states. The summation is over all diabatic microstates of the system. 
If a new phase  is passed by changing the control parameter (such as an external magnetic field acting on a system of spins) from $-\infty$ to $+\infty$ with some rate $\beta$ and if diabatic energies (elements of the diagonal matrix $\hat{B}$ in (\ref{mlz})) of all diabatic states depend linearly on the control parameter then we have $\beta_0-\beta_i = \alpha_i \beta$ for any state $i$, where parameters $\alpha_i$ do not depend on the sweeping rate. The survival probability then satisfies a universal scaling law:
\begin{equation}
{\rm ln}P_{0\rightarrow 0} = -D \beta^{-1},
\label{pt}
\end{equation}
where $D$ is a constant,  $D=2\pi \sum_{i=1}^N |\gamma_{0i}|^2/\alpha_i$. Measurements of this constant by measuring $P_{0\rightarrow 0}$ can reveal a useful information about the coupling of the 
initial ground state to its excitations at the phase transition. 
We would like to stress again that the scaling law (\ref{pt}) is
expected to be truly universal and not restricted by the conditions of
the theory of the adiabatic passage through a quantum critical point
\cite{zurek}. It should be valid for any linear
passage through a region with a phase transition either with or without an exact level crossing point. In mesoscopic systems, such as atomic Bose condensates, the probability $P_{0\rightarrow 0}$ can be made sufficiently large for measurements by increasing the sweeping rate $\beta$.

\section{Conclusions}
We determined state-to-state  transition probabilities in two multistate Landau-Zener models, which have practical relevance for the theory  of dynamic passage through a Feshbach resonance. 
We highlighted the importance of the probability to generate zero number of defects by showing that it can be exponentially enhanced in quantum systems in comparison to classical memoryless stochastic processes and that it shows the universal scaling ${\rm ln}P_{0\rightarrow 0} \sim 1/\beta$ as the function of the sweeping rate $\beta$ through the phase transition. Our solutions also revealed the symmetry of the transition probability matrix that we explored numerically in Appendix for a set of finite chain models. All our tests support the hypothesis that in models of Landau-Zener transitions in chains the transition probability matrix is symmetric.

{\it Acknowledgments.}  Author thanks V. L. Pokrovsky and Ar. Abanov
for useful discussion and M. Anatska for encouragement. Work at LANL was carried out under the auspices of the LDRD/20110189ER and the National Nuclear Security Administration of the U.S. Department of Energy at Los Alamos National Laboratory under Contract No. DE-AC52-06NA25396.

\appendix
\section{Symmetry of transition probability matrix of Landau-Zener models in linear chains}
\begin{figure}[t]
\includegraphics[width=2.4in]{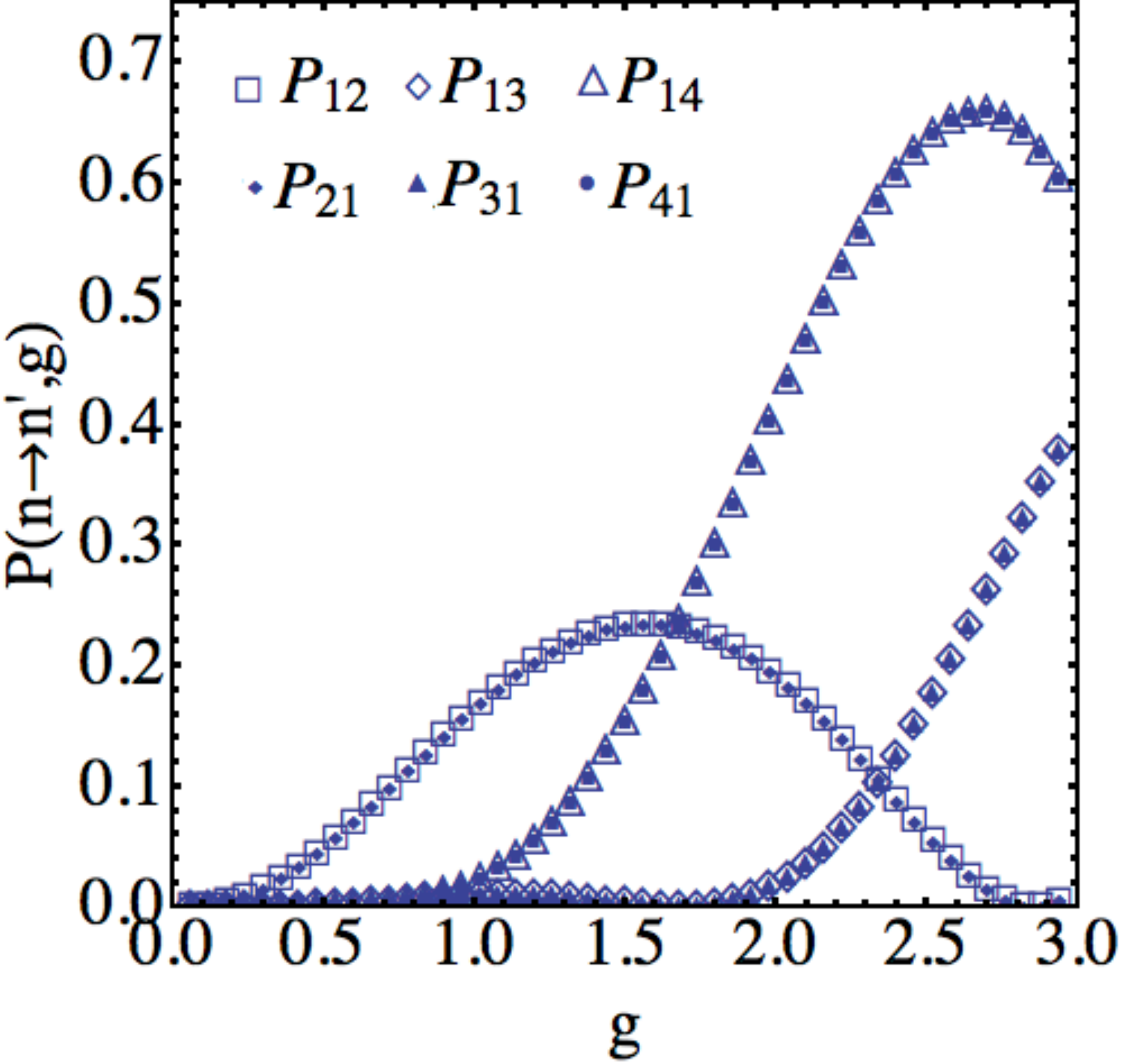}
 \caption{ (Color online) Numerically obtained transition probabilities between the
   diabatic state $n=1$ and other diabatic states in a 4-state MLZM in
   a linear chain. Off-diagonal couplings are proportional to the parameter $g$ but with different, arbitrarily chosen 
 coefficients: $g_{12}=g_{21}=0.32g$,  $g_{23}=g_{32}=0.55g$, $g_{34}=g_{43}=0.7g$. Sweeping rates are $\beta_1-\beta_2=1.1765$, $\beta_2-\beta_3=1.5385$, $\beta_3-\beta_4=0.704$.  Results confirm the symmetry hypothesis. }
 \label{chain4-1d}
 \end{figure}
\begin{figure}[t]
\includegraphics[width=2.4in]{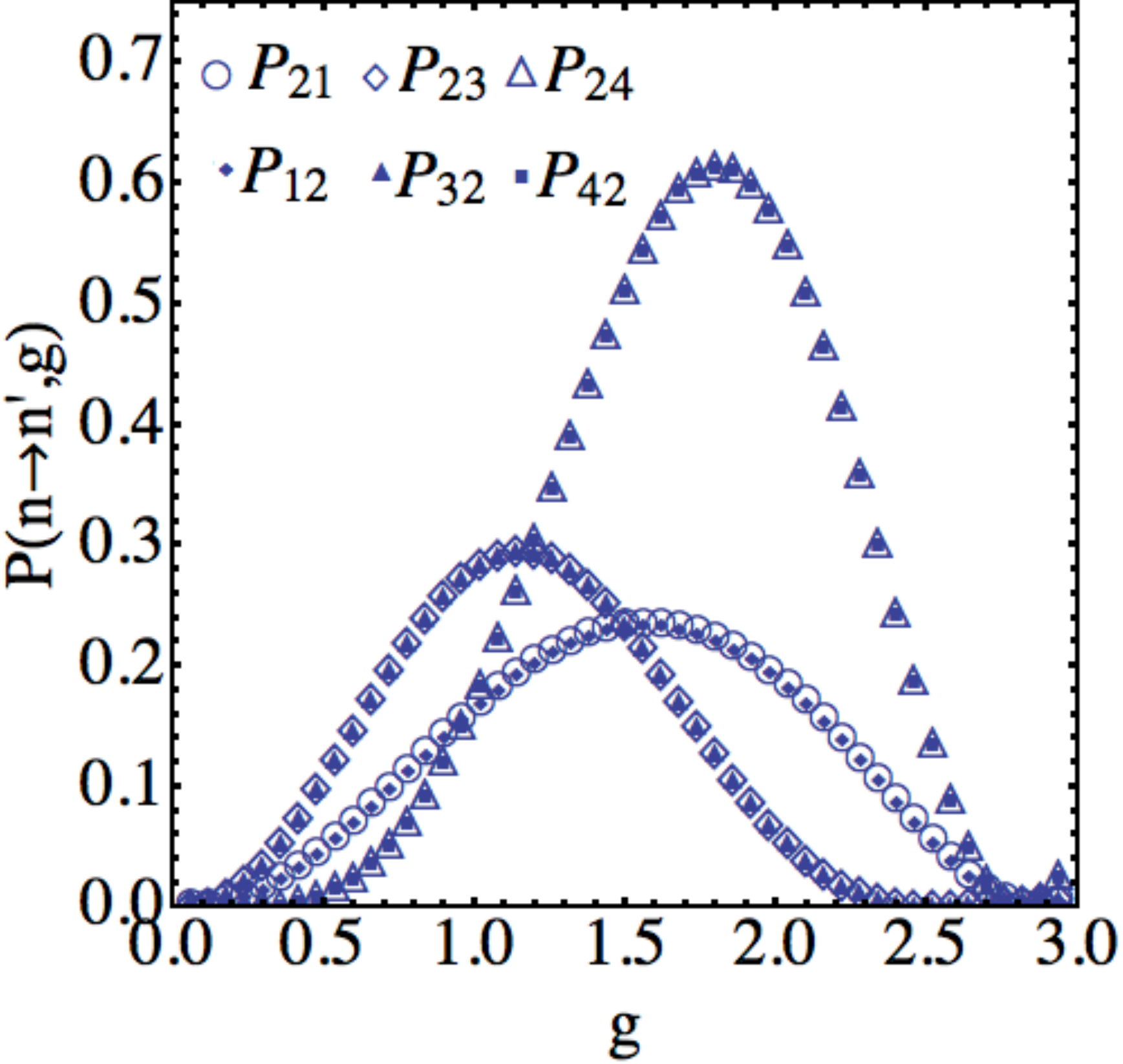}
 \caption{ (Color online) Numerically obtained transition probabilities between the diabatic state $n=2$ and other diabatic states in a 4-state MLZM in a linear chain. Parameters are the same as in Fig.~\ref{chain4-1d}. }
 \label{chain4-2d}
 \end{figure}
\begin{figure}[t]
\includegraphics[width=2.4in]{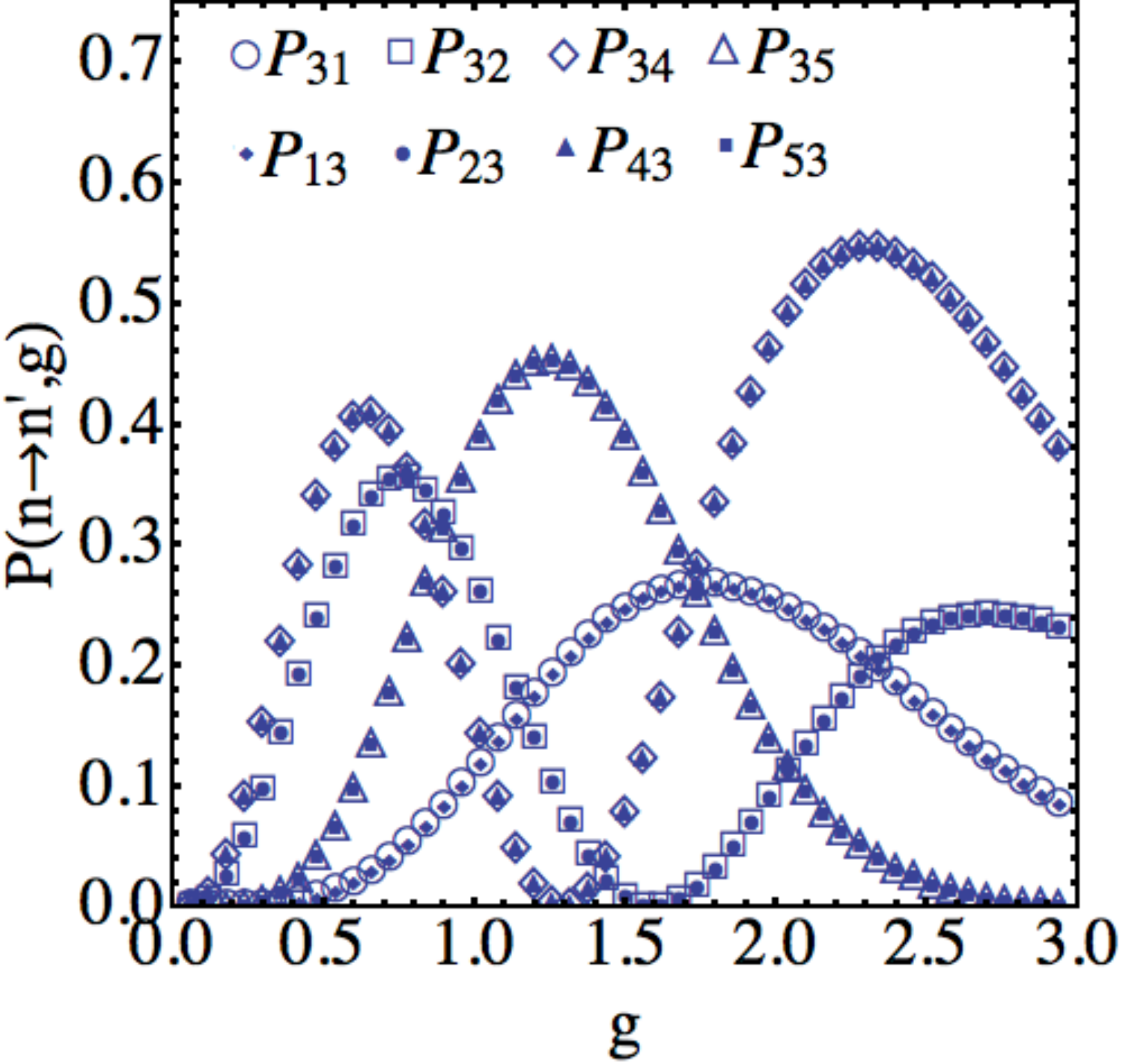}
 \caption{(Color online) Numerically obtained transition probabilities between the diabatic state $n=3$ and other diabatic states in a 5-state MLZM in a linear chain. Off-diagonal couplings are proportional to parameter $g$ but with different, arbitrarily chosen 
 coefficients: $g_{12}=g_{21}=0.32g$,  $g_{23}=g_{32}=0.55g$, $g_{34}=g_{43}=0.7g$, $g_{45}=g_{54}=0.61g$. Sweeping rates are $\beta_n=n$, $n \in (1,2,3,4,5)$. }
 \label{chain5}
 \end{figure}
\begin{figure}[t]  
\includegraphics[width=2.4in]{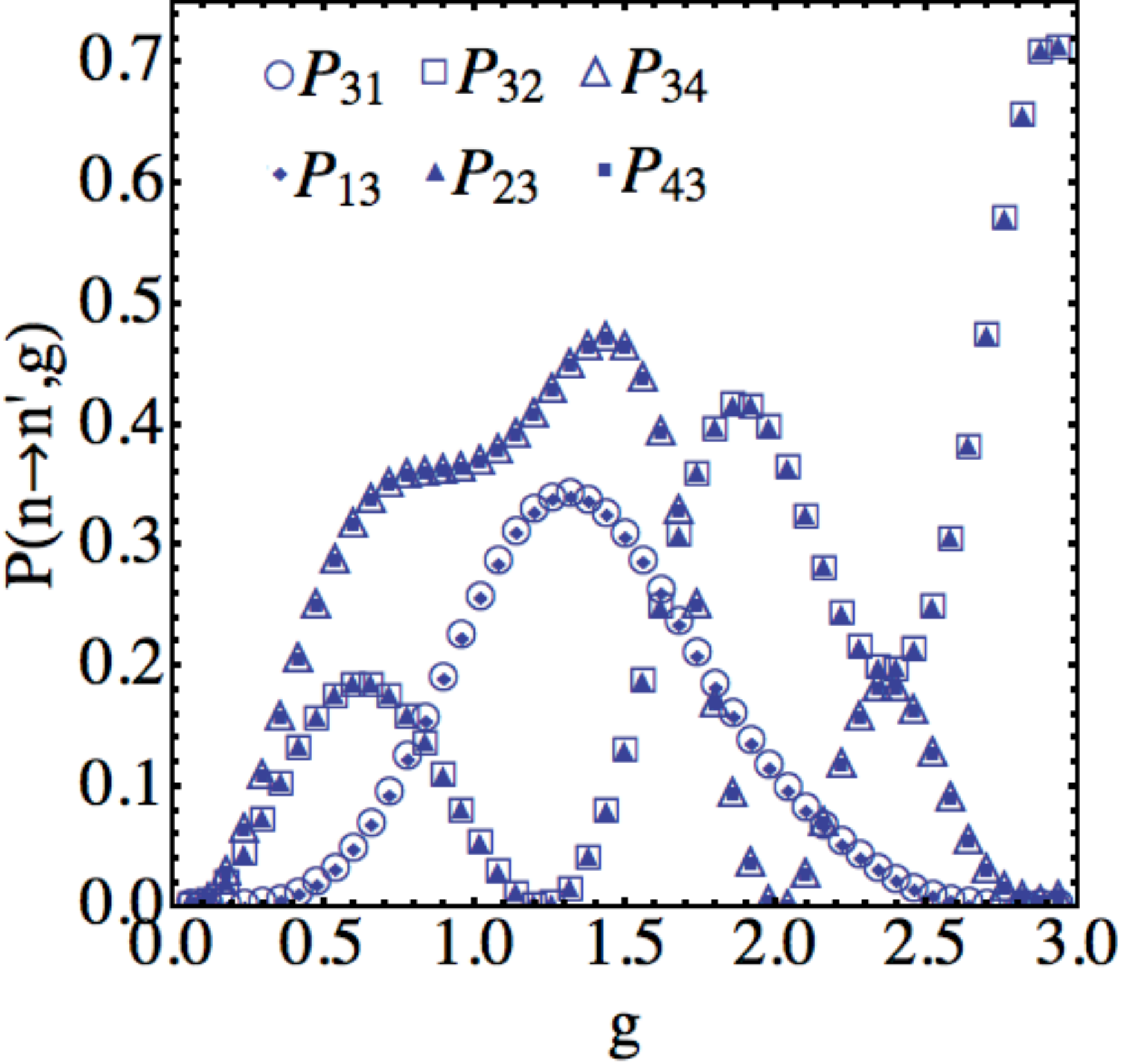}
 \caption{ (Color online) Numerically obtained transition probabilities between the diabatic state $n=3$ and some other diabatic states in a 6-state MLZM. Off-diagonal couplings are proportional to parameter $g$ but with different, arbitrarily chosen 
 coefficients: $g_{12}=g_{21}=0.85g$,  $g_{23}=g_{32}=0.55g$, $g_{34}=g_{43}=0.7g$, $g_{45}=g_{54}=0.92g$, $g_{56}=g_{65}=1.0g$. Sweeping rates are $\beta_1-\beta_2=1.2$, $\beta_2-\beta_3=0.65$, $\beta_3-\beta_4=0.7$,
 $\beta_4-\beta_5=0.91$, $\beta_5-\beta_6=1.0$.    }
 \label{chain6-3d}
 \end{figure}
\begin{figure}[t] 
\includegraphics[width=3.2in]{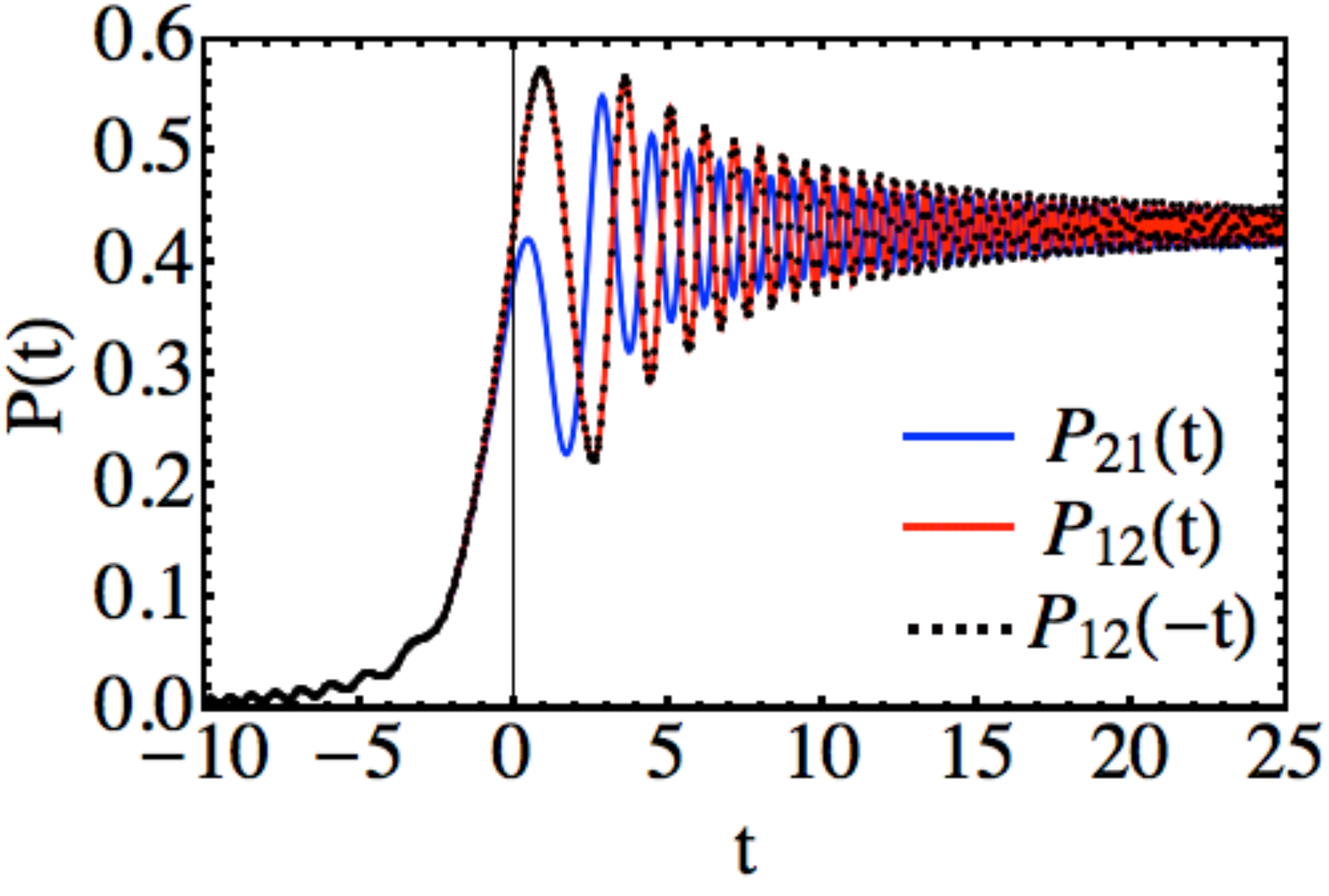}
 \caption{(Color online) Numerically obtained transition probabilities as  functions of time in a 4-state MLZM. {\it Blue curve}: time dependence of the probability of being at the level $n=1$ if the evolution starts at the level $n=2$. Time evolution is from $t=-700$ to $t=700$. {\it Red curve}: time dependence of the probability of being at the level $n=2$ if the evolution starts at the level $n=1$.   {\it Black dots}: The probability of being at the level $n=2$ during the time reversed evolution from $t=700$ backward in time. The evolution  starts at the level $n=1$ at $t=700$. Parameters of the model: $g_{12}=0.82$, $g_{23}=0.55$, $g_{34}=0.7$, $\beta_n=n$, $n \in (1,2,3,4)$. The symmetry $P_{1\rightarrow 2}=P_{2\rightarrow 1}$ is found only asymptotically but at intermediate time, $P_{2\rightarrow 1}(t) \ne P_{1\rightarrow 2}(t)$.
 Time reversed evolution shows that $P_{2\rightarrow 1}(t) \ne P_{1\rightarrow 2}(-t)$  but $P_{1\rightarrow 2}(t) = P_{1\rightarrow 2}(-t)$.  }
 \label{chain4time}
 \end{figure}
 
Models 1 and 2 belong to the class of MLZMs that can be defined as
Landau-Zener transitions in chains. Generally, the evolution for
amplitudes of $N$ states in a chain can be written as
\begin{eqnarray}
\label{nchain} 
i\dot{a}_n &=& \beta_n t a_n +g_{n-1,n} a_{n-1} +g_{n,n+1}^* a_{n+1}, \,\, \\
\nonumber i\dot{a}_1&=&\beta_1 t a_1+g^*_{12}a_2, \quad i\dot{a}_N=\beta_N t a_N+g_{N-1,N}a_{N-1},
\end{eqnarray}
where $n \in (2,\ldots,N-1)$ with some real constants $\beta_k$, $k\in (1,\ldots N)$ and complex constants $g_{k,k+1}$, $k\in (1,\ldots N-1)$ .

Our explicit solutions of Model-1 and Model-2 revealed the symmetry of the transition probability matrix: 
\begin{equation}
P_{n\rightarrow n'}=P_{n' \rightarrow n},
\label{sym}
\end{equation}
which is valid for transitions between any pair of diabatic states $n$ and $n'$. This observation is surprising because the original equations that define our models explicitly break the chiral symmetry. Moreover, from some of the exactly solvable models, such as the Demkov-Osherov model \cite{mlz0}, it is known that generally there is no such a symmetry in the full system (\ref{mlz}). On the other hand, in addition to our models, 
the available solutions of finite size MLZMs in linear chains that include the Majorana's solution for an arbitrary spin \cite{maj} and the exact solution of the 3-state MLZM \cite{mlz5} also show the symmetry  (\ref{sym}).
Hence, based on the available information, we suggest the "symmetry hypothesis" that Eq. (\ref{sym})  is actually asymptotically (i.e. for the time evolution from $-\infty$ to $+\infty$) exact for all MLZMs in linear chains.

Today, the explicit expressions for transition probabilities in a general model (\ref{nchain}) are unknown. Therefore, to test the symmetry hypothesis, we performed a number of numerical tests with models of different complexity. 
Figs.~\ref{chain4-1d},\ref{chain4-2d},\ref{chain5},\ref{chain6-3d} show some of our results for 4-,5-, and 6-state models of the type (\ref{nchain}). Numerical tests with several other systems, including semi-infinite chains, of the type (\ref{nchain}) were also performed but not shown here. We found that {\it all} our numerical tests supported the hypothesis (\ref{sym}) in MLZMs of the type (\ref{nchain}).

The symmetry (\ref{sym}) is {\it not} a simple consequence of some trivial symmetry of (\ref{nchain}). 
An example of a trivial symmetry is the statement that in any system (\ref{nchain}) transition probabilities depend only on absolute values of coupling constants, $|g_{n,n+1}|$. This follows from the fact that a time independent change of the basis $a_n \rightarrow a_ne^{i\phi_n}$ linearly changes phases of $g_{n,n\pm1}$. It is  always possible then to choose $N$ phases $\phi_n$ so that all $N-1$ parameters $g_{n,n+1}$ become real. Another trivial 
fact is that transition probabilities depend only on differences of the level slopes $\beta_i-\beta_j$, which follows from the fact that a uniform shift of all $\beta_k$ by a constant $\beta$, $\beta_k \rightarrow \beta_k+\beta$ for $k\in(1,\ldots N)$, is merely equivalent to the time-dependent phase shift $a_n \rightarrow a_n e^{i\beta t^2/2}$. Such "trivial" symmetries manifest themselves not only asymptotically but also during the full time of the evolution of the system. 
In contrast, Eq. (\ref{sym}) 
is generally {\it not} satisfied at intermediate times during the evolution, as we illustrate in Fig.~\ref{chain4time}. Generally, Eq. (\ref{sym}) is satisfied only in the limit $t \rightarrow +\infty$ when both transition probabilities saturate at the same value. Figure~\ref{chain4time} also shows that in MLZMs in chains the time-reversed process produces the same probability matrix but, alone, this does not explain the asymptotic symmetry (\ref{sym}). 

The symmetry of the probability matrix  considerably reduces the number of unknown parameters. The importance of models of the type (\ref{nchain}) for the theory of the Feshbach resonance has triggered research on developing approximate analytical and numerical approaches to solve such models \cite{gur1,gur2,iti}. The exact symmetry (\ref{sym}) can be  a useful tool to test such approximations and reduce the number of unknown functions during studies of statistics of defects. On the other hand, the existence of such a nontrivial symmetry suggests that there is a way to rigorously prove it. Such a proof should shed new light on the properties of MLZMs and, possibly, it may lead to the expression for the scattering matrix of the whole problem  (\ref{nchain}) in terms of the known special functions.


\end{document}